\documentclass[preprint]{revtex4}
\usepackage{amsfonts}
\usepackage{amssymb,epsfig}
\usepackage{latexsym}
\usepackage{amsmath}
\usepackage{graphicx}

\begin{document}
\title{Reconstruction of modified gravity with ghost dark energy models}
\author{A. Khodam-Mohammadi$^{1,2}$\footnote{Email:\text{khodam@basu.ac.ir}},
M. Malekjani$^{1}$
\footnote{Email:\text{malekjani@basu.ac.ir}}~~and~~ M.
Monshizadeh$^3$ \footnote{Email:\text{monshizadeh470@yahoo.com}}}

\affiliation{$^{1}$ Department of Physics, Faculty of Science,
Bu-Ali Sina University, Hamedan 65178, Iran \\ $^{2}$ Department of
Physics, Faculty of Science, Shiraz University, Shiraz 71454, Iran
\\$^3$ Physics Department, Faculty of Science,
Islamic Azad University, Hamedan branch, Iran}

\begin{abstract}
In this work, we reconstruct the $f(R)$ modified gravity for
different ghost and generalized-ghost dark energy (DE) models in FRW
flat universe, which describe the accelerated expansion of the
universe. The equation of state and deceleration parameter of
reconstructed $f(R)$ gravity have been calculated. The equation of
state and deceleration parameter of reconstructed
$f(R)$-ghost/generalized-ghost DE, have been calculated. We show
that the corresponding $f(R)$ gravity of ghost/generalized-ghost DE
model can behave like phantom or quintessence. Also the transition
between deceleration to acceleration regime is indicated by
deceleration parameter diagram for reconstructed $f(R)$
generalized-ghost DE model.
\end{abstract}

\maketitle


\newpage
\section{Introduction}
The acceleration expansion of the present universe has been accepted
by many observational evidence \cite{allen}. Although the dark
energy (DE) scenario with a negative pressure can drive this
expansion, the modified gravity, so called $f(R)$ gravity, has been
introduced as a self consistent and nontrivial alternative way to
explain early time inflation and present acceleration expansion. In
$f(R)$ gravity, the Einstein-Hilbert action is modified and
generalized by incorporating a new phenomenological function of
Ricci scalar, $f(R)$.

On the other hand the dynamical DE models are proposed in order to
solve DE problems (cosmic coincidence and fine tuning problems).
Recently, the Veneziano ghost DE has been attracted a deal of
attention in the dynamical DE category. The Veneziano ghost is
proposed to solve the $U(1)$ problem in low-energy effective theory
of QCD \cite{witten} and has no contribution in the flat Minkowski
spacetime. In curved spacetime, however, it makes a small energy
density proportional to $\Lambda^3_{QCD}H$, where $\Lambda_{QCD}$ is
QCD mass scale and $H$ is Hubbel parameter. This small vacuum energy
density can be considered as a driver engine for evolution of the
universe. It should be mentioned that the Veneziano ghost DE model
does not violate unitarity, causality, gauge invariance etc
\cite{zhitnitsky, urban, ohta}. In fact the description of DE in
terms of the Veneziano ghost is just a matter of convenience to
describe very complicated infrared dynamics of strongly coupled QCD.
In other words, the veneziano ghost is not a new physical
propagating degree of freedom. The same dynamics can be described
without using the ghost, with some other approaches (e.g. direct
lattice simulations). Also note that this model is totally arisen
from standard model and general relativity. Therefore one needs not
to introduce any new parameter or new degree of freedom and this
fact is the most advantages of ghost DE. With $\Lambda_{QCD}\sim
100Mev$ and $H\sim 10^{-33}ev$, the right order of observed DE
density can be given by ghost DE. This numerical coincidence also
shows that this model gets ride of fine tuning problem \cite{urban,
ohta}. Many authors have already suggested DE model with energy
density as $\rho= \alpha H$ \cite{bjorken,Cai} and $\rho= \alpha
H+\mathcal{O}(H^2)$ \cite{zhitnitsky2}. Generally it is very
difficult to accept such a power like behavior as QCD is a theory
with a mass gap determined by the scale $\sim100$ Mev. The power
like scaling $\sim H$ is due to the complicated topological
structure of strongly coupled QCD, not related to the physical
massive propagating degrees of freedom. In fact the linear in Hubble
constant ``$H$" scaling is not a baseless assumption, but rather has
a strong theoretical support tested in a number of models, and
tested in the lattice QCD simulations \cite{zhitnitsky, urban, ohta,
zhitnitsky2}. Recently the ghost-scalar field models have been
investigated \cite{fernandez}. Further more ghost DE model has been
fitted with current observational data including SNIa, BAO, CMB, BBN
\cite{Cai,Cai2}.

On the other hands, a consistent $f(R)$ function in modified gravity
context, has been reconstructed by many dynamical DE models
\cite{nojiri, karami, setare2}. It is worthwhile to mention that the
modified $f(R)$ gravity model considered as an effective description
of the underlying theory of DE, and considering the ghost DE
scenario as pointing in the direction of the underlying theory of
DE, it is interesting to study how the $f(R)$ gravity can describe
the ghost DE density as effective theory of DE models. This
motivated us to establish the different models of $f(R)$ gravity
according to the ghost and generalized-ghost DE model. In this paper
we want to reconstruct a consistent modified gravity so that it
gives the cosmological evolution of ghost DE model.

\section{General formalism of $F(R)$ gravity}\label{PLECHDE}
The general action of $f(R)$ gravity in Planck mass unit in which
$\hbar=c=8\pi G=1$, is given by \cite{NO2007}
\begin{equation}
S =\int\sqrt{-g}~{\rm d}^4 x\left[\frac{R+f(R)}{2}+\mathcal{L}_{\rm
m}\right],\label{action}
\end{equation}
where $g$, $R$ and $\mathcal{L}_{\rm m}$ are the determinant of
metric $g_{\mu\nu}$, Ricci scalar and matter contribution of the
action, respectively. Here $G$ is gravitational constant and $f(R)$
is an arbitrary function of $R$. In metric formalism, Where the
action is varied only by the metric, by taking the variation of
action with respect to the metric $g_{\mu\nu}$, the following field
equation is obtained
\begin{equation}
\left(R_{\mu\nu}-\frac{1}{2}Rg_{\mu\nu}\right)-
\left[\frac{1}{2}g_{\mu\nu}f(R)-R_{\mu\nu}f'(R)+\big(\nabla_{\mu}\nabla_{\nu}-g_{\mu\nu}\Box\big)f'(R)\right]=
T_{\mu\nu}^{(m)}.\label{field equation}
\end{equation}
Here the prime denotes a derivative with respect to $R$ and
$R_{\mu\nu}$ and $T_{\mu\nu}^{(m)}$ are the Ricci tensor and the
energy-momentum tensor of the matter, respectively.

In a spatially flat-FRW background universe, with following line
element
\begin{equation}
{\rm d}s^2  =  - {\rm d}t^2  + a^2 (t)\left[ {\rm d}r^2+r^2{\rm
d}\theta^2+r^2 \sin^2(\theta){\rm d}\phi^2\right],
\end{equation}
by taking the matter in the prefect fluid form, in which
$T_\nu^{\mu(m)}={\rm diag}(-\rho_m,p_m,p_m,p_m)$, the field equation
(\ref{field equation}) reduces to following modified Friedmann
equations \cite{NO2009,ohta}
\begin{equation}
H^2=\frac{1}{3}\rho_{eff},\label{FiEq1}
\end{equation}
\begin{equation}
2\dot{H}+3H^2=-p_{eff}.\label{FiEq2}
\end{equation}
Here $H=\dot{a}/a$ is the Hubble parameter and
\begin{equation}
\rho_{eff} = \left[-\frac{1}{2}f(R)+3\big(\dot{H}+
H^2\big)f'(R)-18\big(4 H^2\dot{H}+
H\ddot{H}\big)f''(R)\right]+\rho_{m},\label{density}
\end{equation}
\begin{eqnarray}
p_{eff}= &\Big[&\frac{1}{2}f(R)-\big(\dot{H}+3H^2\big)f'(R)
\nonumber\\
&&+6\big(8H^2\dot{H}+6H\ddot{H}+4{\dot{H}}^2+\dddot{H}\big)f''(R)+36
\big(\ddot{H}+4H\dot{H}\big)^2f'''(R)\Big]+p_{m}.\label{pressure}
\end{eqnarray}
The Ricci scalar is obtained as
\begin{equation} R =6\big(\dot{H}+2H^2\big),\label{R}
\end{equation}
where the dot denotes a derivative with respect to cosmic time $t$.

The energy conservation law gives
\begin{equation}
\dot{\rho_{eff}}+3H\rho_{eff}(1+w_{tot})=0,\label{conteq}
\end{equation}
where
\begin{equation}
w_{tot}=\frac{p_{eff}}{\rho_{eff}}
\end{equation}
is the total equation of state (EoS) parameter. In the absence of
any matter field, `$F(R)$ dominated universe', the EoS parameter has
only the gravitational contribution as
\begin{equation}
w_{R}=-1-4
\frac{\dot{H}f'(R)+3\big(3H\ddot{H}-4H^2\dot{H}+4{\dot{H}}^2+\dddot{H}\big)f''(R)
+18\big(\ddot{H}+4H\dot{H}\big)^2f'''(R)} {f(R)-6\big(\dot{H}+
H^2\big)f'(R)+36\big(4 H^2\dot{H}+ H\ddot{H}\big)f''(R)}.\label{WR1}
\end{equation}
In this case the first modified Friedmann equation (\ref{FiEq1})
yields
\begin{equation}
3H^2=\rho_R,\label{FiEq12}
\end{equation}
where $\rho_R$ is the gravitational contribution of energy density
(\ref{density}) for $\rho_m=0$.

Taking time derivative of Eq. (\ref{FiEq12}) and using continuity
equation (\ref{conteq}), one can find a new relation for EoS
parameter as
\begin{equation}
w_R=-1-\frac{2\dot{H}}{3H^2}.\label{WR2}
\end{equation}
For $\dot{H}>0$, we have $w_R<-1$, indicating the phantom DE
dominated universe. In the case of $\dot{H}<0$, we have $w_R>-1$,
representing the quintessence DE dominated universe. The
deceleration parameter in the $F(R)$ dominated universe is obtained
as
\begin{equation}
q_R=-\frac{\ddot{a}a}{\dot{a}^2}=-1-\frac{\dot{H}}{H^2}=\frac{3w_R+1}{2}.\label{q1}
\end{equation}
In the case of $w_R<-1/3$, we have $q<0$ indicating the accelerated
expansion of the universe and for $w_R>-1/3$, we have $q>0$,
representing the decelerated expansion phase.\\
In the theories of modified gravity there are two class of scale
factor that have been usually used in the literature \cite{grav}.
The first one has been given by
\begin{equation}
a(t)=a_0(t_s-t)^{-h},~~~t\leq t_s,~~~h>0.\label{a}
\end{equation}
Using Eqs. (\ref{R}) and (\ref{a}) one can obtain
\begin{equation}
H=\frac{h}{t_s-t}=\left[\frac{hR}{12h+6}\right]^{1/2},~~~\dot{H}=H^2/h,\label{respect
to r},~~~\ddot{H}=\frac{2H^3}{h^2}.
\end{equation}
Here a big rip singularity will be happened at $t=t_s$ \cite{od5}.
It it is easy to see that for the first class of scale factor given
by (\ref{a}) we have $\dot{H}>0$. Hence from (\ref{WR2}), one can
conclude that for this class of scale factor $w_R<-1$, indicating
the phantom DE dominated universe. This scale factor is usually
so-called phantom scale factor.

The second class of scale factor bas been defined as
\begin{equation}
a(t)=a_0t^h,~~~h>0,\label{aQ}
\end{equation}
The Hubble parameter, in this case, is obtained as
\begin{equation}
H=\frac{h}{t}=\left[\frac{hR}{12h-6}\right]^{1/2},~~~\dot{H}=-H^2/h,~~~\ddot{H}=\frac{2H^3}{h^2}.\label{respect
to rQ}
\end{equation}
From  (\ref{WR2}) one can see that $\dot{H}=-H^2/h<0$ represents the
quintessence dominated universe. Hence, the model (\ref{aQ}) is
so-called the quintessence scale factor.\\

In sections (3) and (4), by using the above classes of scale factor,
we reconstruct the different $f(R)$- gravities according to
the ghost and modified ghost DE models.\\
\section{f(R) reconstruction from ghost DE \label{ghost}}
The energy density of ghost DE is given by
\begin{equation}\label{state}
\rho_{\Lambda}=\alpha H
\end{equation}
where $\alpha$ is a constant of model.\\
\subsection{phantom scale factor}
 For the first class of
scale factor (\ref{a}) and using (\ref{respect to r}), the energy
density of ghost DE in (\ref{state}) is written as
\begin{equation}\label{rho2}
\rho_{\Lambda}=\alpha \sqrt{\frac{hR}{12h+6}}
\end{equation}
Equating (\ref{rho2}) and (\ref{density}) by using (\ref{respect to
r}) gives the following solution
\begin{equation}\label{fr1}
f(R)=\lambda_{+}R^{m_{+}}+\lambda_{-}R^{m_{-}}-\alpha \gamma_p
R^{\frac{1}{2}}
\end{equation}
where
\begin{equation}
m_{\pm}=\frac{h+3\pm\sqrt{h^2-10h+1}}{4} \label{mpmp}
\end{equation}
and
\begin{equation}
\gamma_p=\frac{2}{9}\sqrt{\frac{12h+6}{h}}\label{gamap}
\end{equation}
Here $\lambda_{\pm}$ are the constants of integration and can be
obtained from the initial conditions. In order to generating
accelerated expansion at the present time, one can assume that
$f(R)$ is a small constant at the present time as follows
\begin{equation}\label{noji}
f(R_0)=-2R_0,~~~~~and~~~~~f^{\prime}(R_0)\sim 0,
\end{equation}
where $R_0 \sim (10^{33}ev)^2$ is the current curvature
\cite{noji2}. Using the boundary condition (\ref{noji}), the
constants $\lambda_{\pm}$ in (\ref{fr1}) are obtained as
\begin{equation}\label{lam1}
\lambda_{+}=\frac{4\sqrt{R_0}m_{-}+\alpha \gamma_p
(1-2m_{-})}{2R_0^{m_{+}-\frac{1}{2}}(m_{+}-m_{-})}
\end{equation}
and
\begin{equation}\label{lam2}
\lambda_{-}=-\frac{4\sqrt{R_0}m_{+}+\alpha \gamma_p
(1-2m_{+})}{2R_0^{m_{-}-\frac{1}{2}}(m_{+}-m_{-})}
\end{equation}
Inserting (\ref{fr1}) into (\ref{WR1}) and using (\ref{respect to
r}) obtains the EoS parameter of ghost $f(r)$ gravity as
\begin{equation}
w_R=-1-\frac{2}{3h}, \label{wP1}
\end{equation}
which describe the phantom phase of accelerated universe. It must be
mention that, from Eq(\ref{mpmq}), the reconstruction can be
performed provided that
\begin{equation}
0<h\leq (5-2\sqrt{6})~~or~~ h\geq (5+2\sqrt{6})\label{codp}
\end{equation}
The EoS parameter (\ref{wP1}),shows that the reconstructed $f(R)$
gravity from the ghost DE can cross the phantom divide, while in the
scenario of DE, the ghost model can cross the phantom divide if the
interaction between dark matter and DE is included \cite{sheykhi1}.

Using (\ref{wP1}, \ref{q1}), the deceleration parameter is constant
and calculated as
\begin{equation}
q=-1-\frac{1}{h}.
\end{equation}
It shows that, in this case, for all values of $h$ in the range
(\ref{codp}), our universe is expanded under an accelerating phase
$q>0$.

\subsection{Quintessence scale factor}
Now we reconstruct ghost $f(R)$ gravity for the second class scale
factor in (\ref{aQ}). By using (\ref{respect to rQ}), the energy
density of ghost DE in (\ref{state}) can be written as follows
\begin{equation}\label{rho3}
\rho_{\Lambda}=\alpha \sqrt{\frac{hR}{12h-6}}
\end{equation}
Equating (\ref{rho3}) and (\ref{density}) by using (\ref{respect to
rQ}) gives the following solution
\begin{equation}\label{fr2}
f(R)=\lambda_{+}R^{m_{+}}+\lambda_{-}R^{m_{-}}-\alpha \gamma_q
R^{\frac{1}{2}}
\end{equation}
where
\begin{equation}
m_{\pm}=\frac{3-h\pm\sqrt{h^2+10h+1}}{4}\label{mpmq}
\end{equation}
and
\begin{equation}
\gamma_q=\frac{2}{9}\sqrt{\frac{12h-6}{h}}\label{gamaq}
\end{equation}
Using the boundary condition (\ref{noji}), the constants
$\lambda_{\pm}$ in (\ref{fr2}) can be obtained as
\begin{equation}\label{lam11}
\lambda_{+}=\frac{4\sqrt{R_0}m_{-}+\alpha \gamma_q
(1-2m_{-})}{2R_0^{m_{+}-\frac{1}{2}}(m_{+}-m_{-})}
\end{equation}
and
\begin{equation}\label{lam22}
\lambda_{-}=-\frac{4\sqrt{R_0}m_{+}+\alpha \gamma_q
(1-2m_{+})}{2R_0^{m_{-}-\frac{1}{2}}(m_{+}-m_{-})}
\end{equation}
Inserting (\ref{fr2}) into (\ref{WR1}) and using (\ref{respect to
rQ}) results the EoS parameter of ghost $f(r)$ gravity as
\begin{equation}\label{cons2}
w_R=-1+\frac{2}{3h}, ~~~~~~~~~~~~~~~~~ h>1,
\end{equation}
which corresponds to the quintessence regime, i.e., $-1<w_R<-1/3$.
In this case, the deceleration parameter is constant and obtained as
\begin{equation}
q=-1+\frac{1}{h}.\label{qq1}
\end{equation}
In this manner we find that in quintessence regime with $h>1$, the
parameter $q<0$, while for $1/2<h\leq1$, we have $q\geq0$, which
reveal a deceleration expansion at early time. Therefore by this
reconstruction, both deceleration and acceleration have been
permitted, separately for various the parameter $h$.

\section{f(R) reconstruction from generalized-ghost DE}
Here we reconstruct the $f(R)$ gravity for generalized-ghost DE. The
energy density of phenomenological model is given by
\begin{equation}
\rho_{\Lambda}=\alpha H+\beta H^2 \label{GGde}
\end{equation}
where $\alpha$ and $\beta$ are the constants of the model. This
model was first proposed in \cite{Grande} to get an accelerating
universe. For othe motivation to consider this form see
\cite{freese}. When $\beta\rightarrow 0$, we recovered the model
discussed in Sec. \ref{ghost} and for $\alpha\rightarrow 0$, this
model give the holographic DE model with Hubbel scale as IR-cutoff.
Also the subleading term $H^2$ in the ghodt DE model might play a
crucial role in the early evolution of the universe \cite{Cai2}. In
fact in this model we have a `XghostDE' model where `X' denotes the
second term in (\ref{GGde}).

\subsection{Phantom scale factor}
Same as previous section, first we choose the phantom scale factor
(\ref{a}). Therefor the energy density of generalized-ghost DE, by
using (\ref{respect to r}), is rewritten as
\begin{equation}\label{roh6}
\rho_{\Lambda}=\alpha \sqrt{\frac{hR}{12h+6}}+\beta \frac{hR}{12h+6}
\end{equation}
Equating (\ref{roh6}) and (\ref{density}) by using (\ref{respect to
r}) results the following solution for $f(R)$
\begin{equation}\label{frgen1}
f(R)=\lambda_{+}R^{m_{+}}+\lambda_{-}R^{m_{-}}-\alpha \gamma_p
R^{\frac{1}{2}}-\frac{\beta}{3} R
\end{equation}
where parameters $m_{\pm}$ and $\gamma_p$ are the same as Eqs.
(\ref{mpmp}) and (\ref{gamap}), respectively. Applying the boundary
conditions in (\ref{noji}), the constants $\lambda_{\pm}$ in
(\ref{frgen1}) are determined as
\begin{equation}\label{lamdg1}
\lambda_{+}=\frac{2\beta \sqrt{R_0}(1-m_{-})+3\alpha
\gamma_p(1-2m_{-})+12\sqrt{R_0}m_{-}}{6R_0^{m_{+}-\frac{1}{2}}(m_{+}-m_{-})}
\end{equation}
\begin{equation}\label{lamdg2}
\lambda_{-}=-\frac{2\beta \sqrt{R_0}(1-m_{+})+3\alpha
\gamma_p(1-2m_{+})+12\sqrt{R_0}m_{+}}{6R_0^{m_{-}-\frac{1}{2}}(m_{+}-m_{-})}
\end{equation}
Substituting (\ref{frgen1}) into (\ref{WR1}) and using (\ref{respect
to r}) yields the EoS parameter of generalized-ghost $f(R)$ gravity
as
\begin{equation}
w_{R}=-1-\frac{2}{3h}\left(1+\frac{R\beta}{12\alpha H+6\alpha
Hh^{-1}+\beta R}\right).\label{wp2}
 \end{equation}
 The above EoS parameter is time-dependent. Here in contrast with
 constant EoS parameter in (\ref{wP1}), The dynamical behavior of
EoS parameter is given and we see that at infinity it merge to a
constant value (\ref{wP1}) of Sec. \ref{ghost}.

In this case, the time varying deceleration parameter is obtained as
\begin{equation}
q=-1-\frac{1}{h}-\frac{R\beta}{12\alpha hH+6\alpha H+\beta hR}.
\end{equation}
Same as the previous section, this give us the accelerating
expanding universe for all values of $h$ in the range (\ref{codp}).
\subsection{Quintessence scale factor}
We now choose the quintessence scale factor (\ref{aQ}). In this case
the energy density of generalized-ghost DE, by using (\ref{respect
to rQ}), is rewritten as follows
\begin{equation}\label{roh7}
\rho_{\Lambda}=\alpha \sqrt{\frac{hR}{12h-6}}+\beta \frac{hR}{12h-6}
\end{equation}
Equating (\ref{roh7}) and (\ref{density}) by using (\ref{respect to
rQ}) results the following solution for $f(R)$
\begin{equation}\label{frgen2}
f(R)=\lambda_{+}R^{m_{+}}+\lambda_{-}R^{m_{-}}-\alpha \gamma_q
R^{\frac{1}{2}}-\frac{\beta}{3} R
\end{equation}
where parameters $m_{\pm}$ and $\gamma_p$ are the same as Eqs.
(\ref{mpmq}) and (\ref{gamaq}), respectively. Applying the boundary
conditions in (\ref{noji}), the constants $\lambda_{\pm}$ in
(\ref{frgen1}) are determined as
\begin{equation}\label{lamdg22}
\lambda_{+}=\frac{2\beta \sqrt{R_0}(1-m_{-})+3\alpha
\gamma_q(1-2m_{-})+12\sqrt{R_0}m_{-}}{6R_0^{m_{+}-\frac{1}{2}}(m_{+}-m_{-})}
\end{equation}
\begin{equation}\label{lamdg22}
\lambda_{-}=-\frac{2\beta \sqrt{R_0}(1-m_{+})+3\alpha
\gamma_q(1-2m_{+})+12\sqrt{R_0}m_{+}}{6R_0^{m_{-}-\frac{1}{2}}(m_{+}-m_{-})}
\end{equation}
Substituting (\ref{frgen2}) into (\ref{WR1}) and using (\ref{respect
to rQ}) obtains the EoS parameter of generalized-ghost $f(R)$
gravity as
\begin{equation}
w_{R}=-1+\frac{2}{3h}\left(1+\frac{R\beta}{12\alpha H-6\alpha
Hh^{-1}+\beta R}\right),\label{wp3}
 \end{equation}
which can represent the quintessence regime of accelerated universe.
Here we see that the EoS parameter is dynamical, in contrast with
constant EoS parameter in (\ref{cons2}). The dynamical behavior of
EoS parameter is achieved and we see that at infinity it merge to a
constant value (\ref{cons2}) of Sec. \ref{ghost}.

\begin{center}
\begin{figure}[]
\includegraphics[width=9cm]{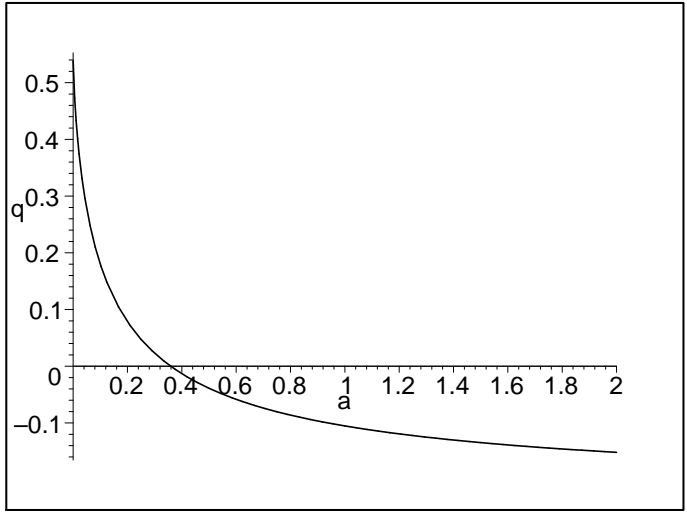}
\caption{The evolution of deceleration parameter, $q$, versus $a$ in
generalized-ghost DE model for parameters:
($h=1.3,~\alpha=2,~\beta=0.3$).} \label{fig1}
\end{figure}
\end{center}

The deceleration parameter, in this case, is time varying and
obtained as
\begin{equation}
q=-1+\frac{1}{h}+\frac{R\beta}{12\alpha h H+6\alpha H+\beta h R}.
\end{equation}
Here in contrast with (\ref{qq1}) in previous section, the
deceleration parameter evolves with time. In Fig. \ref{fig1}, for
$h>1$, an early time transition from the deceleration to
acceleration phases is demonstrated.
\section{conclusion}
The reconstruction of $f(R)$ modified gravity for different ghost
and generalized-ghost DE models in FRW flat universe has been
investigated. The reconstruction was performed by considering two
classes of scale factors, containing i) phantom scale factor,
$a=a_0(t_s-t)^h$ and ii) quintessence scale factor, $a=a_0t^h$. The
equation of state and deceleration parameter of reconstructed
$f(R)$-ghost/generalized-ghost, have been calculated. We showed that
the corresponding $f(R)$ gravity of ghost/generalized ghost DE model
can behave like phantom or quintessence. In $f(R)$-generalized ghost
case, the EoS parameter can vary with time. Also, the transition
between deceleration to acceleration regime is happened at early
time. These behaviors are in contrast with Pure $f(R)$-ghost and
$f(R)$-holographic DE \cite{karami, setare2}. Therefore the
generalized-ghost DE model give us a stronger view of the universe
in $f(R)$ reconstruction point of view.

\end{document}